\documentclass[aps, prb, twocolumn, superscriptaddress, showpacs]{revtex4-1}
\usepackage[colorlinks, linkcolor=blue, anchorcolor=blue, citecolor=blue]{hyperref}
\usepackage{amsmath}
\usepackage{graphicx}
\usepackage{braket}
\usepackage{color}
\usepackage{todonotes}

\begin{document}

\title{On the possibility to detect multipolar order in URu$_2$Si$_2$ by the electric quadrupolar transition of resonant elastic X-ray scattering}

\author{Y. L. Wang}     
\affiliation{Department of Condensed Matter Physics and Materials Science, Brookhaven National Laboratory, Upton, New York 11973, USA} 
\author{G. Fabbris}
\affiliation{Department of Condensed Matter Physics and Materials Science, Brookhaven National Laboratory, Upton, New York 11973, USA} 
\author{D. Meyers} 
\affiliation{Department of Condensed Matter Physics and Materials Science, Brookhaven National Laboratory, Upton, New York 11973, USA} 
\author{N. H. Sung}
\affiliation{Materials Physics and Applications Division, Los Alamos National Laboratory, Los Alamos, New Mexico 87545, USA}
\author{R. E. Baumbach}
\affiliation{Materials Physics and Applications Division, Los Alamos National Laboratory, Los Alamos, New Mexico 87545, USA}
\affiliation{Condensed Matter Group, National High Magnetic Field Laboratory, Florida State University, Tallahassee, Florida 32310, USA}
\author{E. D. Bauer}
\affiliation{Materials Physics and Applications Division, Los Alamos National Laboratory, Los Alamos, New Mexico 87545, USA}
\author{P. J. Ryan} 
\affiliation{Advanced Photon Source, Argonne National Laboratory, Argonne, Illinois 60439, USA}
\affiliation{School of Physical Sciences, Dublin City University, Dublin 9, Ireland}
\author{J.-W. Kim}
\affiliation{Advanced Photon Source, Argonne National Laboratory, Argonne, Illinois 60439, USA}
\author{X. Liu} 
\affiliation{Beijing National Laboratory for Condensed Matter Physics and Institute of Physics, Chinese Academy of Sciences, Beijing 100190, China}
\author{M.€‰ P. €‰M. Dean}
\affiliation{Department of Condensed Matter Physics and Materials Science, Brookhaven National Laboratory, Upton, New York 11973, USA} 
\author{G. Kotliar}
\affiliation{Department of Condensed Matter Physics and Materials Science, Brookhaven National Laboratory, Upton, New York 11973, USA} 
\affiliation{Department of Physics and Astronomy, Rutgers University, Piscataway, New Jersey 08856, USA}
\author{X. Dai} 
\affiliation{Beijing National Laboratory for Condensed Matter Physics and Institute of Physics, Chinese Academy of Sciences, Beijing 100190, China}

\date{\today}

\begin{abstract}
Resonant elastic X-ray scattering is a powerful technique for measuring multipolar order parameters. In this paper, we theoretically and experimentally study the possibility 
of using this technique to detect the proposed multipolar order parameters in URu$_2$Si$_2$ at the U-$L_{3}$ edge with the electric quadrupolar transition. 
Based on an atomic model, we calculate the azimuthal dependence of the quadrupolar transition at the U-$L_{3}$ edge. The results illustrate the potential of this technique 
for distinguishing different multipolar order parameters. We then perform experiments on ultra-clean single crystals of URu$_2$Si$_2$ at the U-$L_{3}$ edge to 
search for the predicted signal, but do not detect any indications of multipolar moments within the experimental uncertainty. 
We theoretically estimate the orders of magnitude of the cross-section and the expected count rate of the quadrupolar transition and compare them to 
the dipolar transitions at the U-$M_4$ and U-$L_3$ edges, clarifying the difficulty in detecting higher order multipolar order parameters in URu$_2$Si$_2$ 
in the current experimental setup.

%% We suggest that possible octupole and hexadecapole order parameters in the hidden order phase of URu$_2$Si$_2$, if present, cannot be detected in the current experiment 
%% due to the very weak signal of quadrupolar transition, which we further justify by theoretically estimating the cross-section and the expected count rate of the 
%% quadrupolar transition and comparing them to the dipolar transitions at the U-$M_4$ and U-$L_3$ edges.  
\end{abstract}


\maketitle

\section{introduction}\label{sec:intro}
The heavy fermion compound URu$_2$Si$_2$ undergoes a phase transition at $T_{HO}=17.5$ K to the so called ``Hidden Order" (HO) phase, 
in which the sharp discontinuous specific heat signals a clear second-order phase transition~\cite{palstra:1985}. 
Earlier studies based on neutron scattering~\cite{broholm:1987,broholm:1991} and muon spin rotation~\cite{maclaughlin:1988} conclude that it is a phase transition to
type-I antiferromagnet (AFM) with the ordered moment polarized along the tetragonal $c$-axis.  
However, the observed ordered moment is anomalously very small ($\sim 0.04\pm 0.01 \mu_{\text{B}}$)~\cite{broholm:1987,broholm:1991,maclaughlin:1988}, which
cannot account for the observed large entropy loss ($\sim 0.2R\text{ln}2$), and the primary order parameter (OP) 
is unlikely to be magnetic dipole. Further high pressure experiments on URu$_2$Si$_2$ find a first-order phase transition from the HO phase to a large moment 
antiferromagnetic (LMAF) phase~\cite{amitsuka:2007,amitsuka:1999,butch:2010}. 
These findings further indicate that the HO phase is distinct from the LMAF and the primary OP should be some complex object which is different from a magnetic dipole.

Theoretically, many different schemes of OPs have been proposed, such as
multipolar order~\cite{santini:1994,santini:1998,ohkawa:1999,santini:2000,kiss:2005,hanzawa:2007,haule:2009,cricchio:2009,harima:2010,kusunose:2011,ikeda:2012},
charge- or spin-density wave~\cite{maple:1986,ikeda:1998,mineev:2005,rau:2012},  
chiral spin state~\cite{gorkov:1992},
orbital antiferromagnetism~\cite{chandra:2002}, 
helicity order~\cite{varma:2006}, 
dynamic symmetry breaking~\cite{elgazzar:2009},
nematic order~\cite{fujimoto:2011}, 
hybridization wave~\cite{dubi:2011},
and hastatic order~\cite{chandra:2013,chandra:2015}. 
However, through 30 years of efforts, there is still a lack of convincing evidence to uncover the HO mystery. 
For a more complete review of the theoretical and experimental progress, see Ref.~\cite{mydosh:2011,mydosh:2014}.

Among the many proposals of OPs, the multipolar order is a promising candidate. Recently, Raman scattering experiments~\cite{buhot:2014,kung:2015,kung:2016} 
find a sharp low energy excitation with $A_{2g}$ symmetry below $T_{HO}$. 
Further analysis~\cite{kung:2015,kung:2016} indicates that this $A_{2g}$ excitation is consistent with the hexadecapolar order proposed by Haule and Kotliar~\cite{haule:2009}.   
However, these Raman scattering experiments provide indirect information about the ground state in the sense that they cannot measure modes at the ordering wavevector. 
Wavevector-resolved techniques are desirable to make more definitive conclusions. Among the many options, resonant elastic X-ray scattering (REXS) is a powerful  
tool to directly detect the order of electrons including complex spin and charge multipoles~\cite{Ament:2011,takeshi:2013}. 
There have been already a few REXS experiments~\cite{isaacs:1990,tatsuya:2005,amitsuka:2010,walker:2011} performed to identify the multipolar order in the HO phase of 
URu$_2$Si$_2$.
Amitsuka \textit{et al.}~\cite{amitsuka:2010} and Walker \textit{et al.}~\cite{walker:2011} performed REXS experiments at the U-$M_{4}$ ($3d_{3/2} \rightarrow 5f$) edge 
below $T_{HO}$ and their results have excluded the possibility of any quadrupolar OPs.
However, the $M_{4}$ edge involves electric dipolar transitions (E1) and has minimal sensitivity to multipoles with a rank larger than 2.
The electric quadrupolar transition (E2) can be used to detect octupole and hexadecapole, but unfortunately, the intensity of E2 is usually much weaker than that of E1.  

Recently, dos Reis \textit{et al.}~\cite{reis:2016} discussed a sizable E2 contribution to the U-$L_{2,3}$ X-ray magnetic circular dichroism (XMCD) signal 
in their study of U compounds. The enhanced E2 signal may be due to the large wavevector $k$ (10.615 $\AA^{-1}$ and 8.699 $\AA^{-1}$) at $L_{2,3}$ edges which means that the term $i\boldsymbol{k} \cdot \boldsymbol{r}$ in the expansion of $e^{i\boldsymbol{k} \cdot \boldsymbol{r}}$ cannot be ignored. 
This unexpected finding provides a promising hope to use the E2 transition to directly detect multipolar OPs in URu$_2$Si$_2$.

In this paper, we theoretically and experimentally study the possibility to detect the proposed multipolar OPs in URu$_2$Si$_2$ via E2 transition of REXS. 
Based on an atomic model, we first calculate the azimuthal dependences to show that it can identify different multipolar OPs by symmetry. 
Then, we do REXS experiments on ultra-clean sample of URu$_2$Si$_2$ single crystal to search for the possible signal of multipolar OPs. 
Finally, we justify the experimental results by theoretically estimating the relative strength of E2 transition at $L_{3}$ edge ($L_{3}$-$E2$) 
compared with E1 transition at $M_{4}$ edge ($M_{4}$-$E1$) and $L_{3}$ edge ($2p_{3/2}\rightarrow 6d, L_{3}$-$E1$), and the expected flux of the scattered photons.

\begin{figure}
\includegraphics[width=0.48\textwidth]{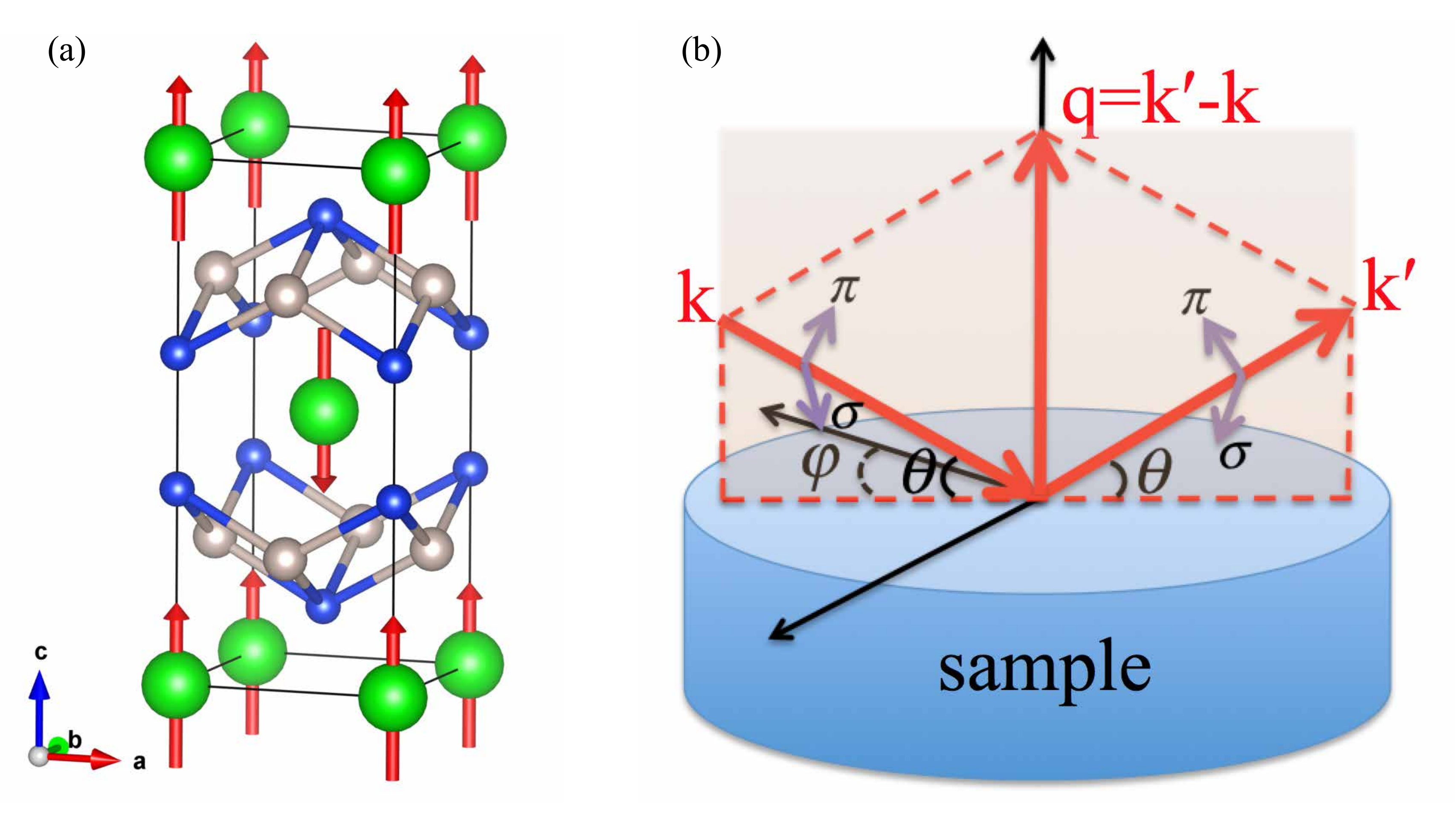}
\caption{(Color online). (a) The crystal structure of URu$_2$Si$_2$. We assume a type-I antiferro-multipolar order on Uranium sites. 
         (b) Illustration of experimental setup. A beam of polarized x-rays $\boldsymbol{k}$ is incident on the $[0 0 1]$ sample face with 
         an angle $\theta$ and scattered by electrons, and then the scattered x-rays 
         $\boldsymbol{k}^{\prime}$ with outgoing angle $\theta$ and specific polarization is analyzed. $\varphi$ is the azimuthal angle. For linear polarization,
         $\pi$ ($\sigma$) polarization is parallel (normal) to the scattering plane.}
\label{fig:setup}
\end{figure}

\section{methods}\label{sec:method}
\subsection{Atomic calculations}
Fig.~\ref{fig:setup}(a) is the crystal structure of URu$_2$Si$_2$, which has a body-centered tetragonal structure.
In the present study, we assume a type-I antiferro-multipolar order on U sites, where sublattice A: U$(0,0,0)$ and sublattice B: U$(0.5, 0.5, 0.5)$ have opposite signs of the 
expectation value of the multipolar moment. The ordering wavevector is $Q_{\text{AF}}=(0,0,1)$.
Fig.~\ref{fig:setup}(b) is a typical experimental setup of REXS. A beam of polarized X-ray $\boldsymbol{k}$ is incident on the sample 
with an angle $\theta$ and then the scattered X-ray $\boldsymbol{k}^{\prime}$ with outgoing angle $\theta$ 
and specific polarization is analyzed. The double-differential cross section~\cite{Ament:2011} for REXS is,
\begin{equation}
\label{eqn:cross}
\frac{\text{d}^2\sigma}{\text{d}\Omega \text{d} E}=r^{2}_{e}m^{2}\omega_{k^{\prime}}^{3}\omega_{k} 
\left|\mathcal{F}_{gg}\left(\boldsymbol{k},\boldsymbol{k}^{\prime},\hbar\omega_{k},\hbar\omega_{k^{\prime}},\boldsymbol{\epsilon},\boldsymbol{\epsilon}^{\prime}\right)\right|^2,
\end{equation}
where, $r_{e}=e^{2}/(4\pi \epsilon_{0}mc^{2})$ is the classical electron radius, $\mathcal{F}_{gg}$ is the scattering amplitude at zero temperature,
\begin{equation}
\label{eqn:scatter}
\mathcal{F}_{gg}\left(\boldsymbol{k},\boldsymbol{k}^{\prime},\hbar\omega_{k},\hbar\omega_{k^{\prime}},\boldsymbol{\epsilon},\boldsymbol{\epsilon}^{\prime}\right) = 
\sum_{n} \frac{\Braket{g|\hat{\mathcal{D}}^{\prime \dagger}|n}\Braket{n|\hat{\mathcal{D}}|g}}{\hbar\omega_{k}+E_{g}-E_{n}+i\Gamma/2},
\end{equation}
where, $\boldsymbol{k}$ is the incoming light with energy $\hbar \omega_{k}$ and polarization $\boldsymbol{\epsilon}$, 
$\boldsymbol{k}^{\prime}$ is the outgoing light with energy $\hbar \omega_{k^{\prime}}$ and polarization $\boldsymbol{\epsilon}^{\prime}$, 
and $\boldsymbol{q}=\boldsymbol{k^\prime}-\boldsymbol{k}$ is the scattering vector.
$\Ket{g}, E_{g}$ is the ground state and $\Ket{n}, E_{n}$ is the eigenstate of the intermediate Hamiltonian including a core-hole.
$\Gamma$ is the lifetime width of the core hole. For U, $\Gamma\approx 8 $ eV at the $L_{3}$ edge and $\Gamma \approx3.5$ eV at the $M_{4}$ edge.
$\hat{\mathcal{D}}$ and $\hat{\mathcal{D}}^{\prime \dagger}$ are the transition operators for absorption and emission processes,
\begin{eqnarray}
\hat{\mathcal{D}} &=& \boldsymbol{P}^{(m)} \cdot \sum_{\boldsymbol{R}} \hat{\boldsymbol{D}}_{\boldsymbol{R}}^{(m)} \nonumber \\
                  &=& \boldsymbol{P}^{(m)} \cdot \left( \sum_{\boldsymbol{R}} e^{i\boldsymbol{k} \cdot \boldsymbol{R}} \sum_{i} \hat{\boldsymbol{r}}_{\boldsymbol{R},i}^{(m)}\right), \\
\hat{\mathcal{D}}^{\prime \dagger} &=& \boldsymbol{P}^{(m) \prime \star} \cdot \sum_{\boldsymbol{R}} \hat{\boldsymbol{D}}_{\boldsymbol{R}}^{(m) \dagger} \nonumber \\
                                   &=& \boldsymbol{P}^{(m) \prime \star} \cdot \left( \sum_{\boldsymbol{R}} e^{-i\boldsymbol{k^{\prime}} \cdot \boldsymbol{R}}  \sum_{i} \hat{\boldsymbol{r}}_{\boldsymbol{R},i}^{(m) \dagger} \right)
\end{eqnarray}
where, $\boldsymbol{R}$ is the site index, $i$ is the index of electron that is bound to site $\boldsymbol{R}$. 
$\boldsymbol{P}^{(m)}$ is a rank-$m$ tensor for geometry part including polarization and wavevector of photon, 
$\hat{\boldsymbol{r}}^{(m)}$ is a single particle rank-$m$ tensor operator of electron.

For E1-E1 transition,
\begin{eqnarray}
\boldsymbol{P}^{(1)}\cdot \hat{\boldsymbol{r}}^{(1)} &=& \epsilon_{x} \hat{x} + \epsilon_{y} \hat{y} + \epsilon_{z} \hat{z},\\
\boldsymbol{P}^{(1) \prime \star} \cdot \hat{\boldsymbol{r}}^{(1) \dagger} &=& \epsilon_{x}^{\prime \star} \hat{x} + \epsilon_{y}^{\prime \star} \hat{y} + \epsilon_{z}^{\prime \star} \hat{z}.
\end{eqnarray}

For E2-E2 transition~\cite{nagao:2006}, 
\begin{eqnarray}
(\hat{\boldsymbol{r}}^{(2)})_{1} &=& \frac{\sqrt{3}}{2}(\hat{x}^2-\hat{y}^2),\\
(\hat{\boldsymbol{r}}^{(2)})_{2} &=& \frac{1}{2}(3\hat{z}^2-\hat{r}^2),\\
(\hat{\boldsymbol{r}}^{(2)})_{3} &=& \sqrt{3}\hat{y}\hat{z},\\
(\hat{\boldsymbol{r}}^{(2)})_{4} &=& \sqrt{3}\hat{z}\hat{x},\\
(\hat{\boldsymbol{r}}^{(2)})_{5} &=& \sqrt{3}\hat{x}\hat{y},
\end{eqnarray} 
and
\begin{eqnarray}
\boldsymbol{P}^{(2)}_{1} &=& \frac{k}{3} \frac{\sqrt{3}}{2}(\epsilon_{x}\tilde{k}_{x} - \epsilon_{y}\tilde{k}_{y}), \\
\boldsymbol{P}^{(2)}_{2} &=& \frac{k}{3} \frac{1}{2} (2\epsilon_{z}\tilde{k}_{z}-\epsilon_{x}\tilde{k}_{x} - \epsilon_{y}\tilde{k}_{y}), \\
\boldsymbol{P}^{(2)}_{3} &=& \frac{k}{3} \frac{\sqrt{3}}{2} (\epsilon_{y}\tilde{k}_{z} + \epsilon_{z}\tilde{k}_{y}),\\
\boldsymbol{P}^{(2)}_{4} &=& \frac{k}{3} \frac{\sqrt{3}}{2} (\epsilon_{z}\tilde{k}_{x} + \epsilon_{x}\tilde{k}_{z}),\\
\boldsymbol{P}^{(2)}_{5} &=& \frac{k}{3} \frac{\sqrt{3}}{2} (\epsilon_{x}\tilde{k}_{y} + \epsilon_{y}\tilde{k}_{x}).
\end{eqnarray}
where, $k$ and $\tilde{\boldsymbol{k}}$ are the length and direction of the wavevector, respectively. 
We assume the absorption and emission process take place at the same site, then the scattering amplitude can be written as,
\begin{equation}
\mathcal{F}_{gg} \propto \sum_{\boldsymbol{R}} e^{-i\boldsymbol{q}\cdot \boldsymbol{R}} F_{gg}^{\boldsymbol{R}},
\end{equation}
with
\begin{equation}
F_{gg}^{\boldsymbol{R}} = \sum_{n} \frac{\Braket{g|\hat{\mathcal{D}}_{\boldsymbol{R}}^{\dagger}|n}\Braket{n|\hat{\mathcal{D}}_{\boldsymbol{R}}|g}}{\hbar \omega_{\boldsymbol{k}} + E_{g} - E_{n} + i\Gamma /2}.
\end{equation} 
where $\hat{\mathcal{D}}_{\boldsymbol{R}}=\boldsymbol{P}^{(m)} \cdot \hat{\boldsymbol{D}}_{\boldsymbol{R}}^{(m)}$ and $\hat{\mathcal{D}}_{\boldsymbol{R}}^{\dagger}=\boldsymbol{P}^{(m)\prime\star} \cdot \hat{\boldsymbol{D}}_{\boldsymbol{R}}^{(m)\dagger}$.

We further make single atom approximation, i.e., approximating the states $\Ket{g}$ and $\Ket{n}$ as single atomic states.  
Then the total scattering amplitude can be written as the summation of the contributions from two sublattices A and B of U atoms,
\begin{equation}
\mathcal{F}_{gg} \propto \sum_{\boldsymbol{R}_{A}} e^{-i\boldsymbol{q}\cdot\boldsymbol{R}_{A}} F_{gg}^{\boldsymbol{R}_{A}} + \sum_{\boldsymbol{R}_{B}} e^{-i\boldsymbol{q}\cdot\boldsymbol{R}_{B}}F_{gg}^{\boldsymbol{R}_{B}},
\end{equation}
where,
\begin{eqnarray}
F^{\boldsymbol{R}_A}_{gg} &=& \sum_{n} \frac{\Braket{g^{A}|\hat{\mathcal{D}}_{\boldsymbol{R}_{A}}^{\dagger}|n^{A}}\Braket{n^{A}|\hat{\mathcal{D}}_{\boldsymbol{R}_{A}}|g^{A}}}{\hbar\omega_{k} + E_{g^{A}} - E_{n^{A}} + i\Gamma/2}, \\
F^{\boldsymbol{R}_B}_{gg} &=& \sum_{n} \frac{\Braket{g^{B}|\hat{\mathcal{D}}_{\boldsymbol{R}_{B}}^{\dagger}|n^{B}}\Braket{n^{B}|\hat{\mathcal{D}}_{\boldsymbol{R}_{B}}|g^{B}}}{\hbar\omega_{k} + E_{g^{B}} - E_{n^{B}} + i\Gamma/2}.
\end{eqnarray}
$\Ket{g^{A}} (\Ket{n^{A}})$ and $\Ket{g^{B}} (\Ket{n^{B}})$ are the ground (intermediate) states of U atoms A and B, respectively. 
In calculation, we choose the ground states to induce opposite signs of the expectation value of multipolar moment at the A and B site.

We use the Cowan-Butler-Thole approach~\cite{cowan:1981,bulter:1981,groot:2008,eli:2010} to exactly diagonalize the atomic Hamiltonian for ground and excited configurations and then get the transition matrix.
For URu$_2$Si$_2$, we assume a $5f^{2}$ ground configuration. For $M_{4}$-$E1$, $L_{3}$-$E1$ and $L_{3}$-$E2$ transitions, the excited configurations are $3d^{5}5f^{3}$,
$2p^{5}5f^{2}6d^{1}$ and $2p^{5}5f^{3}$, respectively.
The Slater integrals $F^{k}, G^{k}$ and spin-orbit coupling (SOC) $\zeta$ of the valence and core electrons are calculated by the Hartree-Fock (HF) methods in Cowan's code~\cite{cowan:1981}.
Usually, HF will overestimate them, so we rescale $F^{k}, G^{k}$ by 80\% and rescale SOC $\zeta$ by 92\% for $2p$ core hole and 96\% for $3d$ core hole, respectively.
The parameters are listed in the Appendix.

According to Hund's rule coupling, the $5f^{2}$ configuration has a ground state with total angular momentum $J=4$ under $SO(3)$ symmetry. 
With $D_{4h}$ crystalline electric field (CEF) symmetry, these nine ground state will split into five singlets and two doublets~\cite{kusunose:2011,sundermann:2016},
\begin{eqnarray}
\Ket{A_{1g}^{(1)}(\alpha)} &=& \text{cos} \alpha \Ket{0} + \frac{\text{sin} \alpha}{\sqrt{2}}\left(\Ket{4}+\Ket{-4} \right), \\
\Ket{A_{1g}^{(2)}(\alpha)} &=& \text{sin} \alpha \Ket{0} - \frac{\text{cos} \alpha}{\sqrt{2}}\left(\Ket{4}+\Ket{-4} \right), \\
\Ket{A_{2g}} &=& \frac{i}{\sqrt{2}}\left(\Ket{4} - \Ket{-4} \right),\\
\Ket{B_{1g}} &=& \frac{1}{\sqrt{2}}\left(\Ket{2} + \Ket{-2} \right),\\
\Ket{B_{2g}} &=& \frac{i}{\sqrt{2}}\left(\Ket{2} - \Ket{-2} \right),\\
\Ket{E_{g}^{(1)}(\beta)} &=& \text{cos} \beta \Ket{\mp 1} + \text{sin} \beta \Ket{\pm 3}, \\
\Ket{E_{g}^{(2)}(\beta)} &=& \text{sin} \beta \Ket{\mp 1} - \text{cos} \beta \Ket{\pm 3}.
\end{eqnarray}

In Ref.~\cite{kusunose:2011}, the authors list the definition of the multipole up to rank-5. We will follow this definition in the present paper
and only discuss multipole up to rank-4 that can, in principle, be detected via the $E2$ transition.
We build different ground states which will induce dipole, quadrupole, octupole and hexadecapole orders.

The ground state that will induce $A_{2+} (A_{2-})$ order can be constructed by a linear combination of $\Ket{A_{2g}}$ and $\Ket{A_{1g}^{(2)}}$,
\begin{equation}
\label{eqn:g2}
\Ket{g^{A(B)}} = \frac{1}{\sqrt{2}}\left( \Ket{A_{2g}} \pm e^{i\eta} \Ket{A_{1g}^{(2)}(40^{o})} \right), 
\end{equation}
where, we take plus sign for $\Ket{g^{A}}$ and minus sign for $\Ket{g^{B}}$.
Note that the subscript $+ (-)$ in $A_{2+} (A_{2-})$ means time-reversal even (odd).
When $\eta=0$, it will induce a $A_{2+}$ hexadecapolar order $H_{z}^{\alpha}=\frac{\sqrt{35}}{2}\overline{J_{x}J_{y}(J_{x}^{2}-J_{y}^{2})}$, 
while $\eta=\pi/2$, it will induce a $A_{2-}$ dipolar order $J_{z}$ and octupolar order $T_{z}^{\alpha}=\frac{1}{2}\overline{J_{z}(5J_{z}^{2}-3J^{2})}$.
This scheme is proposed by Haule and Kotliar~\cite{haule:2009} by a LDA+DMFT calculation. 
In their LDA+DMFT calculation, they also figure out $\alpha$ in $A_{1g}^{(2)}$ should be about $40^{\circ}$. 
$H_{z}^{\alpha}$ is proposed to be the primary OP in the HO phase.
It can be also induced as a secondary OP in the hastatic order scheme~\cite{chandra:2015}. 

The ground state that will induce $B_{1+} (B_{1-})$ order can be written as,
\begin{equation}
\label{eqn:g3}
\Ket{g^{A(B)}}=\frac{1}{\sqrt{2}}\left( \Ket{B_{1g}} \pm e^{i\eta} \Ket{A_{1g}^{(2)}(40^{o})} \right),
\end{equation}
When $\eta=0$, it will induce a $B_{1+}$ quadrupolar order~\cite{ohkawa:1999} $O_{22}=\frac{\sqrt{3}}{2}\overline{J_{x}^{2}-J_{y}^{2}}$ and hexadecapolar order 
$O_{42}=\frac{\sqrt{5}}{4}\overline{(J_{x}^{2}-J_{y}^{2})(7J_{z}^{2}-J^{2})}$, while $\eta=\pi/2$ it will induce a $B_{1-}$ octupolar order~\cite{kiss:2005} 
$T_{xyz}=\sqrt{15}\overline{J_{x}J_{y}J_{z}}$.

The ground state that will induce $B_{2+} (B_{2-})$ order can be written as,
\begin{equation}
\label{eqn:g4}
\Ket{g^{A(B)}}=\frac{1}{\sqrt{2}}\left( \Ket{B_{2g}} \pm e^{i\eta} \Ket{A_{1g}^{(2)}(40^{o})} \right),
\end{equation} 
when $\eta=0$ it will induce a $B_{2+}$ quadrupolar order~\cite{ohkawa:1999} $O_{xy}=\sqrt{3}\overline{J_{x}J_{y}}$ and hexadecapolar order 
$H_{z}^{\beta}=\frac{\sqrt{5}}{2}\overline{J_{x}J_{y}(7J_{z}^{2}-J^{2})}$, while $\eta=\pi/2$, it will induce a $B_{2-}$ octupolar order~\cite{kiss:2005}
$T_{z}^{\beta}=\frac{\sqrt{15}}{2}\overline{J_{z}(J_{x}^{2}-J_{y}^{2})}$.

\subsection{REXS Experiment}
URu$_2$Si$_2$ samples were grown using the Czocharalski method \cite{Matsuda2008}. The residual resistivity ratio (RRR) was measured in various pieces of sample; 
the REXS experiment was performed on the sample with highest RRR (= 361). REXS measurements were performed across the U $L_3$ edge ($\approx$ 17.21~keV) at the 
6-ID-B beamline of the Advanced Photon Source at Argonne National Laboratory. The sample was glued to a Cu holder using GE varnish. 
The holder was placed inside a Be dome filled with He gas, which in turn was mounted on the cold finger of a He closed cycle cryostat. 
A six circle diffractometer was used to move through reciprocal space. Measurements were performed using a scintillator point detector with $1 \times 1$ mm$^2$ slits. 
Tetragonal notation with $a = b = 4.108$~\AA\ and $c = 9.514$~\AA\ is used throughout the paper.

\section{results and discussion}\label{sec:result}
\subsection{Azimuthal dependence for different multipolar order parameters}
\begin{figure*}
\includegraphics[width=1.0\textwidth]{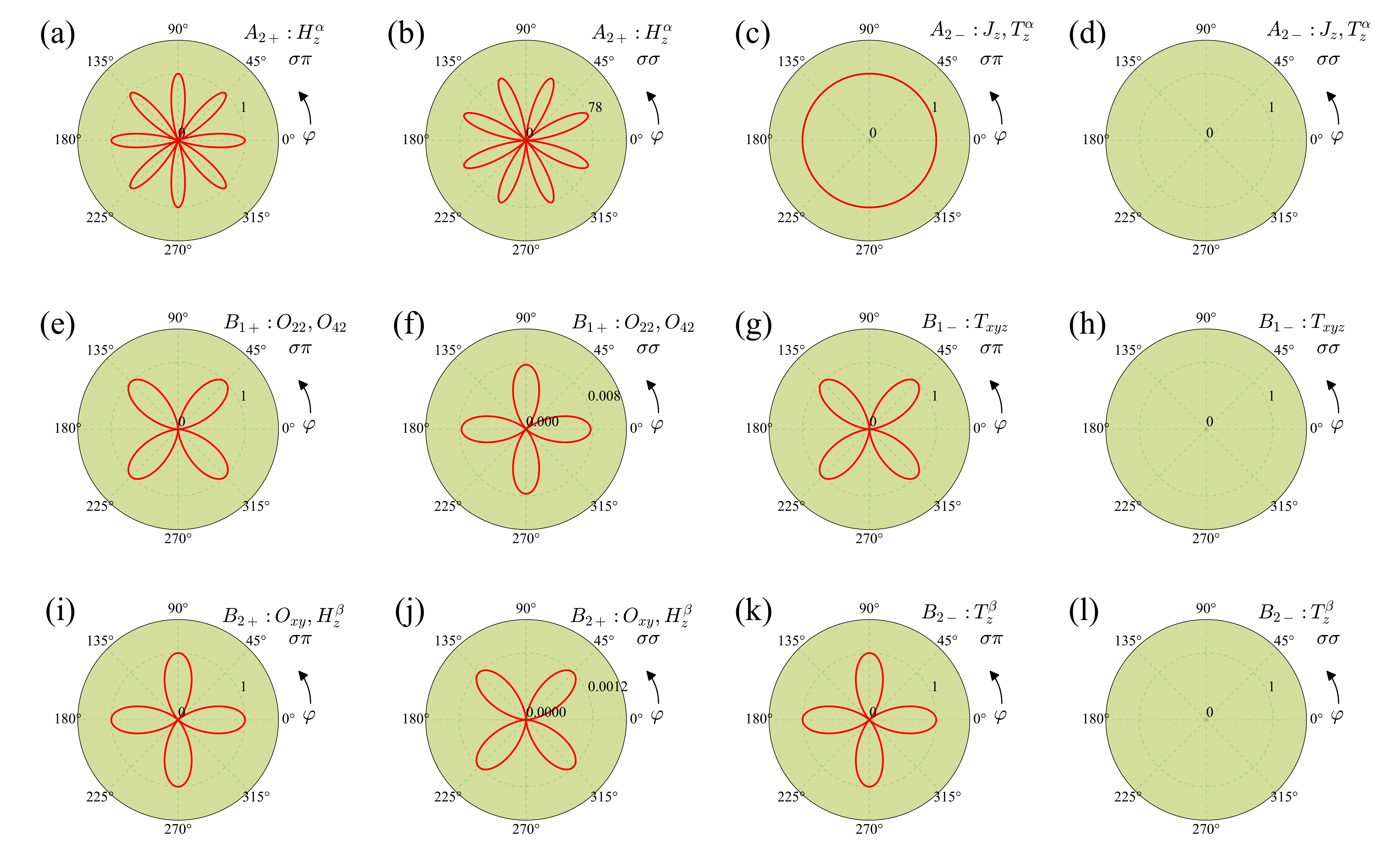}
\caption{(Color online). The calculated azimuthal dependence of a $(0,0,3)$ reflection of $L_{3}$-$E2$ transition  
         in both $\sigma\pi$ and $\sigma\sigma$ channels for different proposals of multipolar OPs. The incident photon energy is 17.167 keV and
         the azimuthal angle is defined with respect to $[100]$ direction. 
         For each proposal, the intensity is normalized by the maximum intensity of its $\sigma\pi$ channel.
         (a,b) $A_{2+}$ hexadecapole $H_{z}^{\alpha}$,
         (c,d) $A_{2-}$ dipole $J_{z}$ and ocutpole $T_{z}^{\alpha}$,
         (e,f) $B_{1+}$ quadrupole $O_{22}$ and hexadecapole $O_{42}$,
         (g,h) $B_{1-}$ octupole $T_{xyz}$,
         (i,j) $B_{2+}$ quadrupole $O_{xy}$ and hexadecapole $H_{z}^{\beta}$,
         (k,l) $B_{2-}$ octupole $T_{z}^{\beta}$.
         }
\label{fig:azi_003}
\end{figure*}

In REXS experiment, azimuthal measurements are used to identify the symmetry of the underlying OPs.
Although Nagao $\textit{et al.}$~\cite{nagao:2006} have figured out the analytic formula of the azimuthal dependences for $E2$ transition,
we still explicitly calculate and plot the azimuthal dependences to show the symmetry difference for different multipolar OPs. 
The results for a $(0,0,3)$ reflection are plotted in Fig.~\ref{fig:azi_003}.
For each multipolar OP, both $\sigma\pi$ and $\sigma\sigma$ channels are plotted, and their intensity is normalized by the maximum of the $\sigma\pi$ channel. 
Fig.~\ref{fig:azi_003}(a,b) plots the results of the $A_{2+}$ hexadecapole $H_{z}^{\alpha}$. 
It shows an eight-fold symmetry with a $\pi/8$ phase shift between the $\sigma\pi$ and $\sigma\sigma$ channels. 
The peak intensity of the $\sigma\sigma$ channel is about 2 orders of magnitude larger than that of the $\sigma\pi$ channel.
The eight-fold symmetry is a characteristic of this $H_{z}^{\alpha}$ hexadecapolar OP.   
Fig.~\ref{fig:azi_003}(c,d) shows the results of the $A_{2-}$ dipole $J_{z}$ and octupole $T_{z}^{\alpha}$. 
It shows a nonzero constant in the $\sigma\pi$ channel and no signal in the $\sigma\sigma$ channel. 
Fig.~\ref{fig:azi_003}(e,f) displays the results of the $B_{1+}$ quadrupole $O_{22}$ and hexadecapole $O_{42}$. 
It shows a $d_{xy}$ wave pattern in the $\sigma\pi$ channel and $d_{x^2-y^2}$ wave pattern in the $\sigma\sigma$ channel.
In Fig.~\ref{fig:azi_003}(g,h) the results of $B_{1-}$ octupole $T_{xyz}$ shows a $d_{xy}$ wave pattern in the $\sigma\pi$ channel and no signal in the $\sigma\sigma$ channel. 
Fig.~\ref{fig:azi_003}(i,j) plots the results of $B_{2+}$ quadrupole $O_{xy}$ and hexadecapole $H_{z}^{\beta}$ exhibiting a $d_{x^2-y^2}$ wave pattern in the $\sigma\pi$ channel and $d_{xy}$ pattern in the $\sigma\sigma$ channel. Finally, the results of the $B_{2-}$ octupole $T_{z}^{\beta}$ are shown in 
Fig.~\ref{fig:azi_003}(k,l) where a $d_{x^2-y^2}$ wave pattern is seen in the $\sigma\pi$ channel and nothing is seen in the $\sigma\sigma$ channel. In general, we find that there are no signals in $\sigma\sigma$ channel for time-reversal broken OPs.
The azimuthal dependence show different symmetries for different multipole, so it can be used to distinguish multipolar OPs.

\subsection{REXS Results}
\begin{figure}
\includegraphics{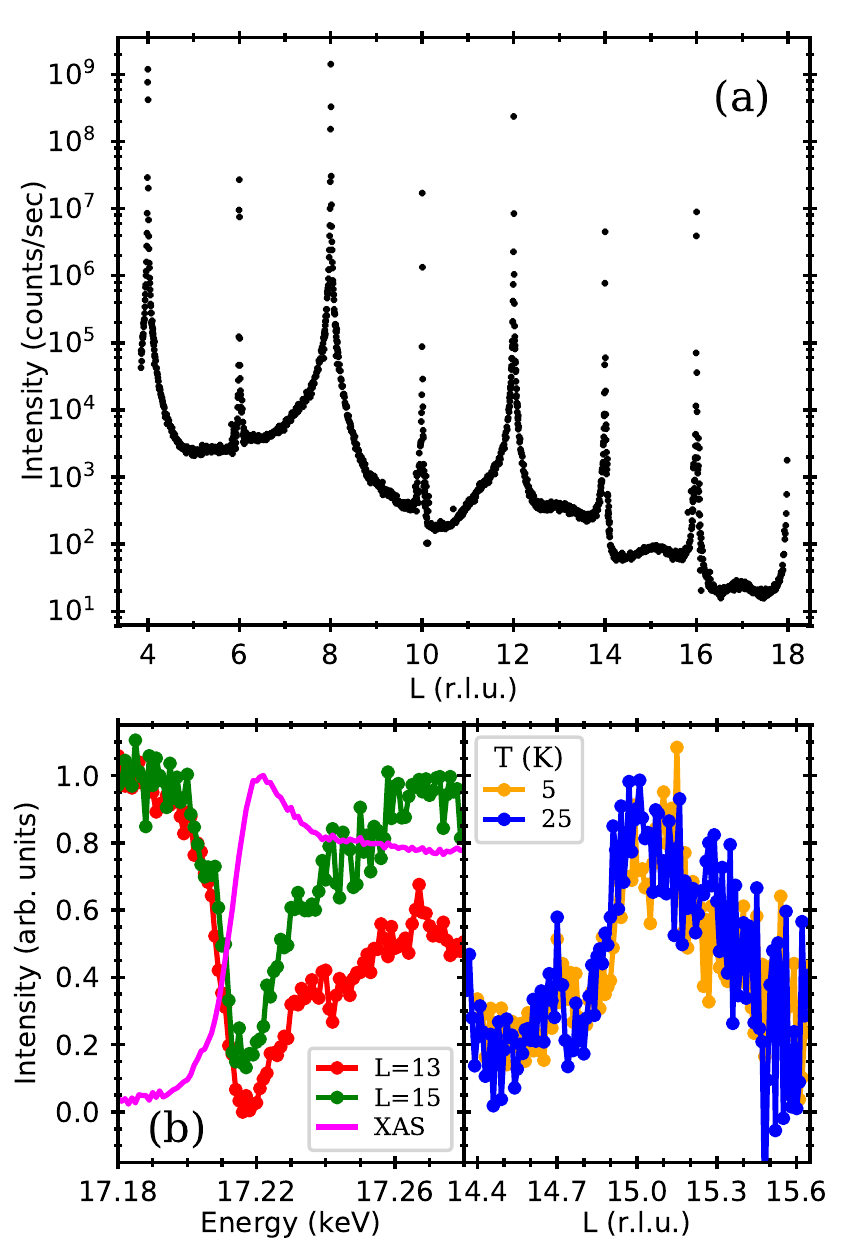}
\caption{(a) X-ray diffraction $L$ dependence measured along $(0,0,L)$ direction using 17.215 keV and $\varphi = 1.2^{\circ}$. (b) Energy dependence of (0,0,13) 
and (0,0,15) Bragg peaks together with the U $L_3$ edge XANES. (c) Temperature dependence of (0,0,15) Bragg peak.}
\label{exp}
\end{figure}

The body centered tetragonal structure of $\mathrm{URu_2Si_2}$ forbids Bragg peaks with $H + K + L = 2n + 1$. We infer that the HO state breaks the body centered 
symmetry by creating inequivalent U sites, thus allowing Bragg peaks at these once forbidden positions. We performed an extensive search for HO Bragg peaks 
along $(0,0,L)$ and $(1,0,L)$ directions; results for the former are displayed in Fig.~\ref{exp}. Broad  peaks are observed at $(0,0,2n+1)$. However, these peaks persist 
through the phase transition at $T_{HO}$, strongly suggesting that these are not related to the HO phase. Additionally, no resonance enhancement is observed across 
the U $L_3$ edge. This suggests that the HO is not accessible through the $E1$ or $E2$ transitions using experiments of this type. These results are consistent with former studies~\cite{amitsuka:2010,walker:2011} in which no quadrupolar OPs are found. 
However, we still cannot exclude the possibility of octupole and hexadecapole due to the weak signal of the $E2$ transition. 

Despite our negative result in the search for the HO, additional experiments are needed to definitely prove the existence (or absence) of the octupole or hexadecapole OPs. 
Designing experimental techniques to enhance the sensitivity to the $E2$ transition at U-$L_{3}$ edge is needed to observe higher rank multipoles.
One of such techniques is the Borrmann spectroscopy~\cite{batterman:1964,pettifer:2008}. The Borrmann effect refers to the anomalous transmission of X-rays
through very perfect single-crystal slabs when they are in symmetric Laue diffraction condition~\cite{batterman:1964}.  This effect can be interpreted by the theory of dynamical diffraction of X-rays~\cite{batterman:1964}. It is a consequence of multiple coherent interference of the incident and diffracted 
beams which produces a total electric field with almost zero amplitude but largely enhanced gradient at the crystal planes. 
The dipolar transition is thus suppressed because it is proportional to the amplitude of the electric field and, on the contrary, the quadrupolar transition 
will be largely enhanced because it is proportional to the gradient of the electric field. Therefore, we may have a chance to detect strong quadrupolar signal, 
for example at U-$L_{3}$ edge. 
In Ref.~\onlinecite{pettifer:2008}, Pettifer \textit{et al.} indeed observed very strong quadrupolar peak in the absorption spectrum at $L_{1}$, 
$L_{2}$ and $L_{3}$ edges of Gadolinium in a $4f$ compound gadolinium gallium garnet. However, no results of $5f$ compounds have been reported, so it is worth to try
in $5f$ compounds, such as URu$_2$Si$_2$. 
Borrmann spectroscopy requires samples that are much thicker than the nominal X-ray penetration depth and sufficiently perfect that at least 
some x-rays can transmit through the sample without encountering defects, which may be a challenge for sample growth.

Polarization analysis of the outgoing X-rays can also be advantageous (despite the strong reduction in X-ray throughput that it imposes) because, as we will demonstrate, 
the HO Bragg peak should be observed in the $\sigma\pi$ channel. 
Additionally, identifying the energy and cross-section of the $L_{3}$- $E2$ transition will greatly facilitate the search for superlattice peaks.

\begin{figure*}
\includegraphics[width=0.95\textwidth]{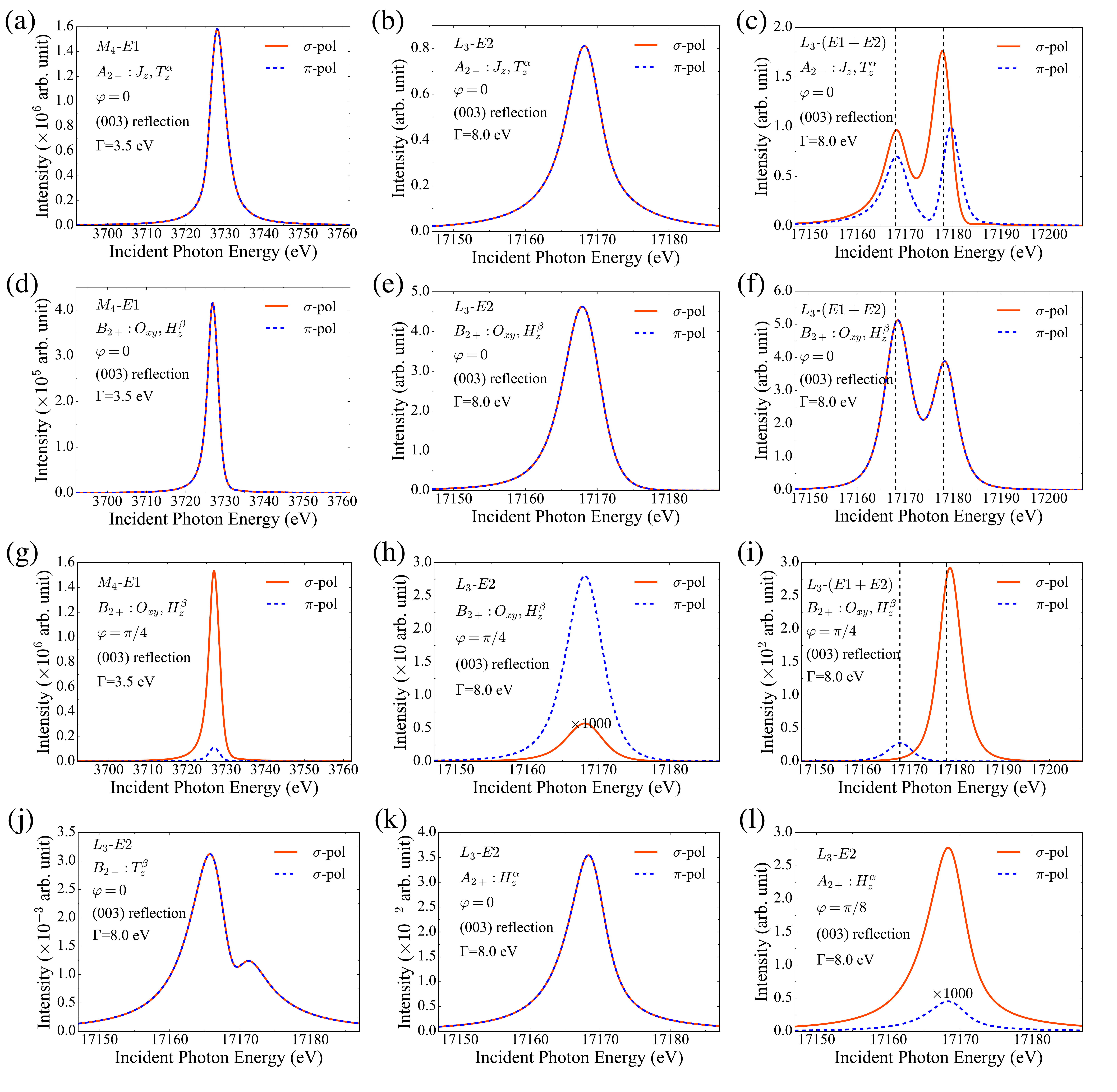}
\caption{(Color online). $(0,0,3)$ REXS intensity as a function of incident photon energy and polarization. 
The incoming light is linearly polarized and the polarization of the outgoing light is not analyzed. We compare the results of the $M_{4}$-$E1$ transition with the
$L_{3}$-$E2$ transition and the $L_{3}$-($E1$ + $E2$) transition.
We consider different ordering schemes:
(a,b,c) Antiferro-dipole $J_{z}$      and antiferro-octupole $T_{z}^{\alpha}$ at $\varphi=0$.  
(d,e,f) Antiferro-quadrupole $O_{xy}$ and antiferro-hexadecapole $H_{z}^{\beta}$ at $\varphi=0$.
(g,h,i) Antiferro-quadrupole $O_{xy}$ and antiferro-hexadecapole $H_{z}^{\beta}$ at $\varphi=\frac{\pi}{4}$. 
(j) Antiferro-octupole $T_{z}^{\beta}$ at $\varphi=0$. (k) Antiferro-hexadecapole $H_{z}^{\alpha}$ at $\varphi=0$. 
(l) Antiferro-hexadecapole $H_{z}^{\alpha}$ at $\varphi=\pi/8$.}
\label{fig:strength}
\end{figure*}

\subsection{Intensity estimation of the $L_{3}$-$E2$ transition}
We further justify the negative experimental results by estimating the intensity of the $L_{3}$-$E2$ transition.
Usually, the intensity of the $E2$ transition will be much weaker than that of the $E1$ transition. This is mainly caused by the very small overlap integral of $r^{2}$ 
between the core hole and valence orbitals. Thus, it is critical to give an estimation of the relative intensity of the $L_{3}$-$E2$ transition compared with
known experiments which have strong intensity, such as the $M_{4}$-$E1$ transition.
Roughly, the relative intensity between $L_{3}$-$E2$ and  $M_{4}$-$E1$ is,
\begin{equation}
\frac{I(L_{3}-E2)}{I(M_{4}-E1)} \propto \left(\frac{k}{3} \frac{\omega_{L_{3}}}{\omega_{M_{4}}} \frac{\Braket{2p|r^{2}|5f}}{\Braket{3d|r|5f}}\right)^{4} \left(\frac{\Gamma_{M_{4}}}{\Gamma_{L_{3}}}\right)^2 ,
\end{equation}
and that between $L_{3}$-$E2$ and $L_{3}$-$E1$ is,
\begin{equation}
\frac{I(L_{3}-E2)}{I(L_{3}-E1)} \propto \left(\frac{k}{3} \frac{\Braket{2p|r^{2}|5f}}{\Braket{2p|r|6d}}\right)^{4},
\end{equation} 
where, $\omega_{L_{3}}$ and $\omega_{M_{4}}$ are the X-ray frequency of the $L_{3}$ and $M_{4}$ edges, their ratio is about 4.6. Based on the HF calculations, the overlap integral ratio
$\Braket{2p|r^2|5f}/\Braket{3d|r|5f} \approx 0.013$ and $\Braket{2p|r^2|5f}/\Braket{2p|r|6d} \approx 0.2$,  respectively. $\Gamma_{M_{4}}$ and $\Gamma_{L_{3}}$ are the core-hole
lifetime width for $M_{4}$ and $L_{3}$ edge, respectively, and their ratio is $\Gamma_{M_{4}}/\Gamma_{L_{3}} \approx 0.4$. For the $L_{3}$ edge, $k/3 \approx 2.9$. 
Thus, the intensity of $L_{3}$-$E2$ is about $10^{-4}$ times smaller than that of $M_{4}$-$E1$ and $10^{-1}$ times smaller than that of $L_{3}$-$E1$.  
Here, we should note that $6d$ orbitals are much broader in URu$_2$Si$_2$, which will lead to larger overlap integrals than those based on the atomic $6d$ orbitals, so $L_{3}$-$E2$ is not just one order of magnitude smaller than that of $L_{3}$-$E1$. 
We may expect larger overlap integrals for the $M_{3}$-$E2$ ($3p_{3/2} \rightarrow 5f$) transition, so we also calculate the relative intensity between $M_{3}$-$E2$  
and $M_{4}$-$E1$. The results show that the intensity of $M_{3}$-$E2$ is also about $10^{-4}$ times smaller than that of $M_{4}$-$E1$. The reason is that,
although the calculated overlap integral $\Braket{3p|r^2|5f}$ is about 14 times larger than that of $\Braket{2p|r^2|5f}$, both of the X-ray frequency and wavevector of 
$M_{3}$ edge is about 0.25 times smaller than that of $L_{3}$ edge, as a result, the enhancement effect from the larger overlap integral is cancelled out. 
The intensity of $M_{3}$-$E2$ is not stronger than that of $L_{3}$-$E2$.  

However, this rough estimation does not consider many details of the scattering process, such as the ground state and the intermediate excited states, the interference effects of
intermediate states, the smearing effect of core-hole lifetime width and the geometry of the experimental setup. 
To give a better estimation, we exactly diagonalize the atomic ground and excited Hamiltonians to get the eigenstates and the transition matrix, 
and then we choose different ground states and experimental geometries to calculate the cross section according to Eqn.~\ref{eqn:cross} and Eqn.~\ref{eqn:scatter}.

The calculated results of a $(0,0,3)$ reflection are shown in Fig.~\ref{fig:strength}. The azimuthal angle $\varphi$ is defined with respect to the $[100]$ direction and
the polarization of outgoing light is not analyzed. We plot both the $\sigma$ and $\pi$ polarizations of the incident light. 
The difference of energy levels between $6d$ and $5f$ is set to be 10~eV. We assume a type-I antiferro-multipolar order with $Q_{AF}=(0,0,1)$ in the simulation.
Fig.~\ref{fig:strength}(a,b,c) are the results for the ground state (Eqn.~\ref{eqn:g2}) that induces $A_{2-}$ orders: dipole $J_{z}$ and octupole $T_{z}^{\alpha}$. The 
$E1$ transition can only detect $J_{z}$ but $E2$ can detect both of them. The azimuthal angle is set to be $\varphi=0$. 
For this ground state, the intensity of $L_{3}$-$E2$ is about $10^{-6}$ times smaller than that of $M_{4}$-$E1$. 
However, the intensity of $L_{3}$-$E2$ is almost the same order of magnitude as that of $L_{3}$-$E1$ transition. 
In Fig.~\ref{fig:strength}(c), the left peak is from the $E2$ transition and the right peak is from the $E1$ transition.
Fig.~\ref{fig:strength}(d,e,f,g,h,i) plots the results for the ground state (Eqn.~\ref{eqn:g4}) that induces $B_{2+}$ order: quadrupole $O_{xy}$ and hexadecapole $H_{z}^{\beta}$. 
In Fig.~\ref{fig:strength}(d,e,f) the azimuthal angle is $\varphi=0$. We find that the intensity of $L_{3}$-$E2$ transition is $10^{-5}$ times smaller than that of
$M_{4}$-$E1$ transition and has the same order of magnitude as that of $L_{3}$-$E1$.
In Fig.~\ref{fig:strength}(g,h,i), the azimuthal angle is set to be $\varphi=\pi/4$. For $\sigma$ polarization, the intensity of $L_{3}$-$E2$ is about $10^{-9}$ 
times smaller than that of $M_{4}$-$E1$ and $10^{-5}$ smaller than that of $L_{3}$-$E1$. 
However, for $\pi$ polarization, it is only $10^{-5}$ times smaller than that of $M_{4}$-$E1$ and much larger than that of $L_{3}$-$E1$ so that there is only a $E2$ peak.
Fig.~\ref{fig:strength}(j) is the result for the ground state (Eqn.~\ref{eqn:g4}) that induces $B_{2-}$ octupolar order $T_{z}^{\beta}$. 
The intensity is at least 8 order of magnitude smaller than that of $M_{4}$-$E1$. 
Another $B_{1-}$ octupole $T_{xyz}$ has the same order of magnitude as that of $T_{z}^{\beta}$. 
Fig.~\ref{fig:strength}(k,l) are the results for the ground state (Eqn.~\ref{eqn:g2}) that induces the $A_{2+}$ hexadecapolar order $H_{z}^{\alpha}$. 
For $\varphi=0$, both $\sigma$ and $\pi$ polarizations are at least 7 order of magnitude smaller than that of $M_{4}$-$E1$.
For $\varphi=\pi/8$, $\sigma$ polarization is about 5 order of magnitude smaller than that of $M_{4}$-$E1$.
We emphasize that the atomic calculation underestimate the intensity of $L_{3}$-$E1$ transition due to the itinerant character of $6d$ orbitals, 
so the intensity of $L_{3}$-$E1$ should be much larger than that of $L_{3}$-$E2$ in reality.  

Based on these atomic results, we find that there are many factors that will affect the REXS cross-section, such as the interference of the intermediate states, 
the interference effect of core-hole lifetime width, the experimental geometry and the details of the ground states. 
Overall, the intensity of $L_{3}$-$E2$ transition is at least 5 or 6 orders of magnitude smaller than that of $M_{4}$-$E1$, so the signal of $L_{3}$-$E2$ transition 
is indeed very weak compared with $M_{4}$-$E1$. We also note that the $5f$ electrons are not completely localized and they have partial 
itinerant character in URu$_2$Si$_2$, which leads to the importance of the band effects in the REXS cross-section. To account for these effects, the combination of 
 more advanced first-principle calculations, such as density functional theory plus dynamical mean-field theory (DFT+DMFT), with REXS cross-section calculations is needed. Despite this, the simple atomic simulations still give us preliminary estimations about the strength of the $E2$ transition. 

To further confirm the weakness of $L_{3}$-$E2$ signal, we estimate the flux of the scattered photons by calculating the absolute value of the cross-section.
For a typical flux of $10^{11} \text{ph}/\text{s}/100 \text{meV}/(100\times 100 \mu m^{2})$, a rough upper bound of the flux of scattered photon is 
$10^{4} \text{ph}/\text{s}/\text{eV}/\text{rad}$ for $M_{4}$-$E1$ transition, while it is $10^{-1}\sim 10^{-2} \text{ph}/\text{s}/\text{eV}/\text{rad}$ for $L_{3}$-$E2$ transition. 
This makes it very difficult to be detected in experiments, which is consistent with the experimental results.   

%% summary
\section{summary}\label{sec:sum}
In summary, we have studied the possibility to detect multipolar OPs in URu$_2$Si$_2$ by REXS in the U $L_{3}$-$E2$ transition channel.
The REXS experiments do not find any clear signal indicating multipolar OPs. 
An estimation based on atomic calculations indicates that the intensity of the $L_{3}$-$E2$ transition is indeed much smaller than that of $M_{4}$-$E1$ transition 
and the flux of the scattered photons is too small such that it is very difficult to detect the $E2$ signal. 
It seems that it is still not practical to use the $E2$ transition of currently available REXS experiment to detect the multipolar OPs.
Developing experimental techniques to enhance $E2$ signal is urgently needed to identify the multipolar OPs not only in URu$_2$Si$_2$ but also in other compounds, 
such as UO$_2$, NpO$_2$ and Ce$_{1-x}$La$_{x}$B$_{6}$~\cite{santini_rmp:2009}.

%% acknowledge
\section{acknowledgments}\label{sec:ack}
We thank Frank de Groot for valuable discussions. 
This work was supported by the U.S. Department of energy, Office of Science, Basic Energy Sciences as a part of the Computational Materials Science Program through
the Center for Computational Design of Functional Strongly Correlated Materials and Theoretical Spectroscopy.
G.F.\ and D.M.\ were supported by the U.S. Department of Energy, Office of Science, Office of Basic Energy Sciences, under Contract No. DE-SC00112704, and Early Career Award Program under Award No. 1047478. X.L. is supported by MOST (Grant No.2015CB921302) and CAS (Grant No. XDB07020200). This research used resources of the Advanced Photon Source, a U.S. Department of Energy (DOE) Office of Science User Facility operated for the DOE Office of Science by Argonne National Laboratory under Contract No. DE-AC02-06CH11357.
Work at Los Alamos National Laboratory was performed under the auspices of the U.S. Department of Energy, Office of Basic Energy Sciences, Division of Materials Sciences and 
Engineering.
\appendix
\section{Slater integrals and spin-orbit coupling parameters}
\begin{table}[htp]
\caption{Slater integrals and spin-orbit coupling parameters for ground configuration $5f^{2}$ (in eV). \label{tab:gs}}
\centering
\begin{tabular}{ccccc}
\hline
\hline
 $F^{0}_{ff}$  & $F^{2}_{ff}$  & $F^{4}_{ff}$   & $F^{6}_{ff}$   & $\zeta_{5f}$ \\
\hline
 0.291         & 7.611         & 4.979          & 3.655          & 0.261       \\
\hline
\hline
\end{tabular}
\end{table}

\begin{table}[htp]
\caption{Slater integrals and spin-orbit coupling parameters for excited configuration $2p^{5}5f^{3}$ (in eV). \label{tab:l3e2_ex}}
\centering
\begin{tabular}{cccccccc}
\hline
\hline
 $F^{0}_{ff}$  & $F^{2}_{ff}$  & $F^{4}_{ff}$   & $F^{6}_{ff}$ & $F^{0}_{pf}$ & $F^{2}_{pf}$ & $G^{2}_{pf}$ & $G^{4}_{pf}$ \\
\hline
 0.306         & 7.984         & 5.232          & 3.845        & 0.005        & 0.497        & 0.082        & 0.053 \\
\hline
$\zeta_{5f}$ & $\zeta_{2p}$ \\
\hline
0.302        & 2517.292 \\
\hline
\hline
\end{tabular}
\end{table}

\begin{table}[htp]
\caption{Slater integrals and spin-orbit coupling parameters for excited configuration $2p^{5}5f^{2}6d^{1}$ (in eV). \label{tab:l3e1_ex}}
\centering
\begin{tabular}{cccccccc}
\hline
\hline
 $F^{0}_{ff}$  & $F^{2}_{ff}$  & $F^{4}_{ff}$   & $F^{6}_{ff}$ & $F^{0}_{pf}$ & $F^{2}_{pf}$ & $G^{2}_{pf}$ & $G^{4}_{pf}$ \\
\hline
 0.307         & 8.278         & 5.447          & 4.011        & 0.102        & 0.528        & 0.087        & 0.056        \\
\hline
$F^{0}_{pd}$   & $F^{2}_{pd}$  & $G^{1}_{pd}$   & $G^{3}_{pd}$ & $F^{0}_{fd}$ & $F^{2}_{fd}$ & $F^{4}_{fd}$ & $G^{1}_{fd}$ \\
\hline
0.022          & 0.272         & 0.238          & 0.142        & 0.139        & 3.750        & 2.050        &  1.938       \\
\hline
$G^{3}_{fd}$   & $G^{5}_{fd}$  & $\zeta_{5f}$   & $\zeta_{6d}$ & $\zeta_{2p}$ \\
\hline
1.562          & 1.213         & 0.321          & 0.435        & 2517.236  \\
\hline
\hline
\end{tabular}
\end{table}

\begin{table}[htp]
\caption{Slater integrals and spin-orbit coupling parameters for excited configuration $3d^{5}5f^{3}$ (in eV). \label{tab:m4e1_ex}}
\centering
\begin{tabular}{ccccccc}
\hline
\hline
 $F^{0}_{ff}$  & $F^{2}_{ff}$  & $F^{4}_{ff}$   & $F^{6}_{ff}$ & $F^{0}_{df}$ & $F^{2}_{df}$ & $F^{4}_{df}$ \\
\hline
 0.307         & 8.020         & 5.258          & 3.865        & 0.102        & 2.051        & 0.952        \\
\hline
$G^{1}_{df}$   & $G^{3}_{df}$ & $G^{5}_{df}$ & $\zeta_{5f}$ & $\zeta_{3d}$ \\
\hline
1.602          & 0.969        & 0.678        & 0.301        & 70.449 \\
\hline
\hline
\end{tabular}
\end{table}

%% references
\bibliography{URu2Si2}

\end{document}